# The Age of the Milky Way Inner Halo

Jason S. Kalirai[1,2]

jkalirai@stsci.edu

[1]Space Telescope Science Institute, 3700 San Martin Drive, Baltimore, MD, 21218

[2]Center for Astrophysical Sciences, Johns Hopkins University, Baltimore, MD 21218

The Milky Way galaxy is observed to have multiple components with distinct properties, such as the bulge, disk, and halo.  Unraveling the assembly history of these populations provides a powerful test to the theory of galaxy formation and evolution, but is often restricted due to difficulties in measuring accurate stellar ages for low mass, hydrogen-burning stars.[1,2]  Unlike these progenitors, the "cinders" of stellar evolution, white dwarf stars[3], are remarkably simple objects and their fundamental properties can be measured with little ambiguity from spectroscopy[4,5].  Here I report observations and analysis of *newly* formed white dwarf stars in the halo of the Milky Way, and a comparison to published analysis of white dwarfs in the well-studied 12.5 billion-year-old globular cluster Messier 4.  From this, I measure the mass distribution of the remnants and invert the stellar evolution process to develop a new relation that links this final stellar mass to the mass of their immediate progenitors, and therefore to the age of the parent population.  By applying this technique to a small sample of four nearby and kinematically-confirmed halo white dwarfs, I measure the age of local field halo stars to be 11.4 +/- 0.7 billion years.  This age is directly tied to the globular cluster age scale, on which the oldest clusters formed 13.5 billion years ago.  Future (spectroscopic) **observations of newly formed white dwarfs in the Milky Way halo can be used to reduce the present uncertainty, and to probe relative differences between the formation time of the *last* clusters and the inner halo.**

Figure 1(a) illustrates the successful discovery of almost 2000 white dwarfs in the nearest globular star cluster, Messier 4. The deep Hubble Space Telescope (*HST*) ACS observations were recently obtained (Guest Observer Program 10146 – PI L. Bedin)[6], and analyzed using new methods[7]. The stellar evolution process that produces these remnants runs like clockwork, since all of the Messier 4 stars formed at the same time – 12.5 +/- 0.5 billion years ago[8]. The brightest objects at *F606W* = 22.5 – 23.5 represent the newly formed remnants of progenitor hydrogen-burning stars that just exhausted their nuclear fuel. The mass of the progenitor stars can be calculated accurately from stellar evolution models since the age of the population is well measured[9]. For Messier 4, this mass is $M_{INITIAL}$ = 0.802 (+0.007; -0.011) $M_{SUN}$. The fainter stars on the Messier 4 white dwarf cooling sequence, at *F606W* = 28 – 29, are "older white dwarfs" that evolved from more massive progenitors earlier in the star cluster's history.

The rich white dwarf cooling sequence of the globular cluster Messier 4 offers a rare opportunity to anchor a new relation that links the final mass of stellar evolution to the age of the population. In Figure 1(b – g), I present published analysis of Keck Telescope multi-object spectroscopy of a half dozen newly formed white dwarfs in Messier 4 (collected from 2005 - 2008; PI. R. M. Rich)[10]. The composition of white dwarfs is simple; a carbon/oxygen core at high pressure surrounded by a helium mantle and a thin atmosphere of hydrogen[11]. Unlike A-dwarfs, the hydrogen atom Balmer lines are strongly pressure broadened. We reproduce these observed profiles with the latest white dwarf atmosphere models, which include updated Stark broadening calculations of the hydrogen atom[12]. The fundamental parameters of each star, including the temperature, gravity, and mass, are measured through a simultaneous fit to both low and high order Balmer lines. These results indicate that the mass of white dwarfs forming today in Messier 4 is $M_{FINAL}$ = 0.529 +/- 0.012 $M_{SUN}$. This is in excellent agreement with both theoretical predictions of the masses of white dwarfs forming today in globular clusters[13], and with an independent (but indirect) measurement in the 12.5 billion year cluster NGC 6752[14].

The absolute calibration of both the initial and final stellar mass at a well-measured (old) age provides the necessary input to calculate the formation time of the Milky Way *field* halo. The only missing ingredient is knowledge of the mass distribution of white dwarfs that are forming *today* in the halo. Previous searches for white dwarfs in the halo have successfully uncovered cool remnants with temperatures of less than 5000 K[15-17]. Such white dwarfs are difficult to date. The cooling rates of the stars depend on their masses, and the masses cannot be measured due to a lack of spectral features at these temperatures. The total age of each star is the combined age of the progenitor lifetime and the (appreciable) white dwarf cooling age.

The spectra of four nearby field white dwarfs is presented in Figure 1(h − k), along with a theoretical fit based on the *same* updated models used to analyze the Messier 4 remnants[12]. These four stars are carefully selected out of 398 white dwarfs in the SPY survey (SN Ia Progenitor Survey), and are measured as kinematic members of the Galactic halo based on three dimensional velocity measurements[18,19]. The temperatures of these stars confirm their nature as newly formed white dwarfs from progenitors that have just exhausted their hydrogen supply.  Using the same method as described above for the Messier 4 white dwarfs, we measure the average mass of the four Milky Way halo white dwarfs to be $M_{FINAL}$ = 0.551 +/- 0.005 $M_{SUN}$.

The mass distribution of the six white dwarfs at the bright tip of the Messier 4 cooling sequence, and the four newly formed white dwarfs in the Galactic halo is shown in Figure 2(a). Through the uniform treatment of both populations, we find that the halo white dwarfs exhibit a larger mass by approximately 4% (0.02 $M_{SUN}$). We interpret this difference to reflect a small difference in the mass of the stellar core of the progenitor star that is presently leaving the hydrogen burning stage in each population.  For such low mass hydrogen burning stars, the post hydrogen-burning evolutionary time scales are virtually identical over small changes in mass, and recent studies have now convincingly demonstrated that progressively lower mass hydrogen burning stars evolve to form lower mass white dwarfs[10,20,21].

To calibrate the measured difference in the mass of white dwarfs forming today in these two populations, we first construct a new relation between the initial and final mass of stars.  The relation is anchored on the Messier 4 measurement of $M_{INITIAL}$ = 0.802 (+0.007; -0.011) $M_{SUN}$ and $M_{FINAL}$ = 0.529 +/- 0.012 $M_{SUN}$.  The $\Delta M$ = 0.02 $M_{SUN}$ difference in the core mass of the star implies an initial mass of the progenitor Milky Way halo stars of 0.825 (+0.009; -0.013) $M_{SUN}$. Next, using the same stellar evolution models that were applied to the globular cluster Messier $4^9$, we calculate the age of the stellar halo near the position of the Sun to be 11.4 +/- 0.7 billion years (Figure 2b). The uncertainty in this measurement derives from the spread in masses of the small sample of four halo white dwarfs, and can be improved considerably by increasing the number of spectroscopically measured remnants in future studies. The relation that links the mass of remnants forming today to the parent population's age is,

$$Log (Age/Gyr) = (Log(M_{FINAL}/M_{SUN} + 0.270) - 0.201)/-0.272.$$

This new relation is directly calibrated on the globular cluster age scale, as defined through a homogenous imaging survey of 60 clusters with $HST$[22]. The entire data set was subjected to a uniform set of stellar evolution models with updated physics[8,9,23]. The results indicate a clear age-metallicity gradient in the population, defined strongly by more than 50 of the clusters. As shown in Figure 2 (b), the Milky Way's most metal-poor clusters formed 13.5 billion years ago and the most metal-rich systems formed 12.0 billion years ago. Only 9 of the clusters are younger than Messier 4. These results therefore suggest that the local Milky Way halo formed 2 billion years after the first globular clusters, approximately at the same time as the last clusters.

Observations of galaxies in the nearby Universe demonstrate that the process of galaxy assembly proceeds in a hierarchical framework. The stellar halo of the Milky Way represents the premier hunting ground to unravel the archaeology of when and how these processes occurred. Recent observations in the Milky Way suggest that the galaxy's halo has populations, distinct in their abundances and kinematics[24,25]. The latest SPH + N-body simulations of the formation of stellar halos are also finding evidence for a dual origin to the halo. In addition to an outer halo that is dominated by accreted stars from satellite disruption, the inner few tens of kiloparsecs of galaxy halos can contain 50% stars that formed in-situ[26,27]. The local white dwarf population in this study is akin to this latter population, and the derived age measurement of 11.4 +/- 0.7 billion years agrees very well with the prediction that 70% of the in-situ population in galaxy formation simulations formed by a redshift of 3 (i.e., 11.5 billion years ago)[26].

Stars that are now in the outer, accreted halo of the Milky Way are predicted to have formed a few billion years before the in-situ star formation of the inner halo[27]. This component is also likely more metal-poor than the [Fe/H] = -1.6 inner halo.[24] Future surveys of newly formed white dwarfs with kinematic characteristics of the outer halo population can test this. For example, if the outer halo was accreted 13.5 billion years ago (consistent with the age of the oldest globular clusters), then we predict the masses of white dwarfs forming today should be 0.51 $M_{SUN}$ or lower depending on how metal-poor the population is.

## Acknowledgements

The data presented in this paper were obtained at the W.M. Keck Observatory, which is operated as a scientific partnership among the California Institute of Technology, the University of California and the National Aeronautics and Space Administration. The Observatory was made possible by the generous financial support of the W.M. Keck Foundation. The author wishes to recognize and acknowledge the very significant cultural role and reverence that the summit of Mauna Kea has always had within the indigenous Hawaiian community. I am most fortunate to have the opportunity to conduct observations from this mountain. Also based on observations obtained at the Paranal Observatory of the European Southern Observatory.

We wish to thank Aaron Dotter, Pierre Bergeron, and Pier-Emmanuel Tremblay for useful discussions related to stellar evolution and ages. We also thank Uli Heber for providing us reduced spectra from the SPY Survey. I wish to acknowledge the team members of the original study of the Messier 4 white dwarfs[10]

## Competing financial interests

The author declares no competing financial interests.

## Corresponding author

Correspondence to: Jason S. Kalirai (jkalirai@stsci.edu)

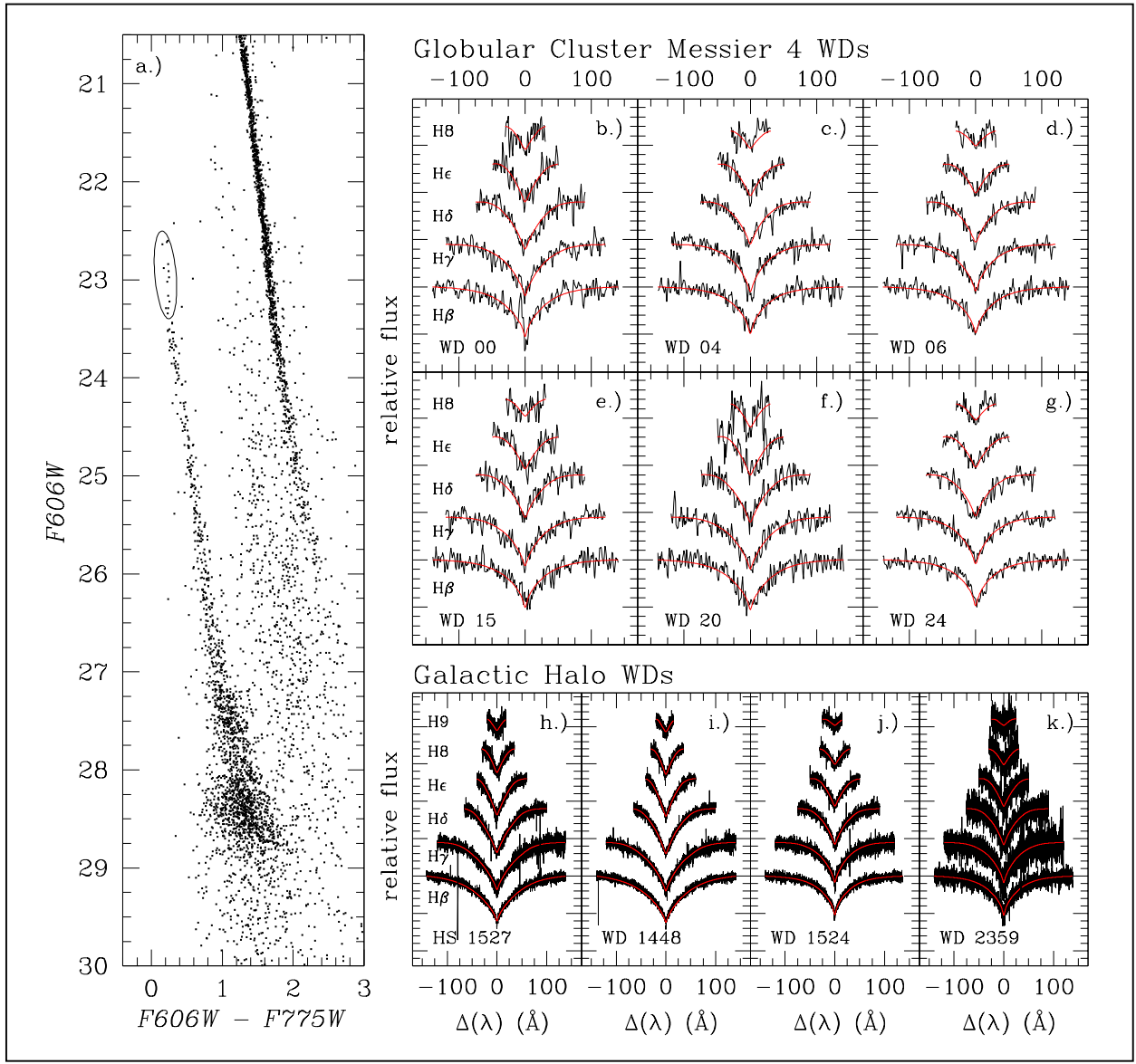

**Figure 1: Spectroscopic Examination of White Dwarfs in Messier 4 and the Milky Way Halo**

**a.)** The stellar cinders of previous generation hydrogen-burning main-sequence stars pile up along the white dwarf cooling sequence of the nearby, 12.5 billion year old globular cluster Messier 4[8]. The brightest of these white dwarfs (within the circle) have been cooling for less than 100 million years, and are therefore the end products of stars that are just today evolving from the hydrogen-burning phase (i.e., stars with $M_{INITIAL} = 0.802$ (+0.007; -0.011) $M_{SUN}$ for Messier 4). **b.) – j.)** Spectroscopic observations of the brightest Messier 4 white dwarfs **(b. – g.)**[10] and the newly formed white dwarfs in the Milky Way halo from the SPY Survey **(h. – j.)**. Specifically, of the seven halo candidates in SPY[19], we accept the three stars that are more than 3.5 sigma outliers from the thin and thick disk distributions (WD 1524, HS 1527, WD 1448) as well as a fourth star that has a clear halo-like orbit (WD 2359). The white dwarf WD 0252 is rejected on these criteria, and we also note that the mass of white dwarf is <0.40 $M_{SUN}$, and so it could not have formed through a normal channel of stellar evolution (e.g., it could be a helium-core white dwarf that was in a binary). The temperatures of the four stars are measured to be 14000 – 20000 K, and so their cooling ages are only 25 – 300 million years. Each panel illustrates all of the broad Balmer absorption lines in a single star that belongs to the respective populations. These pressure-broadened spectral features are reproduced using the latest white dwarf atmosphere models (with updated Stark broadening physics) to reveal the temperatures, gravities, and masses of the stars[12]. To avoid systematic errors from the two different resolutions, the model spectra were convolved to the resolution of the respective data sets (using a Gaussian function) prior to the fitting.

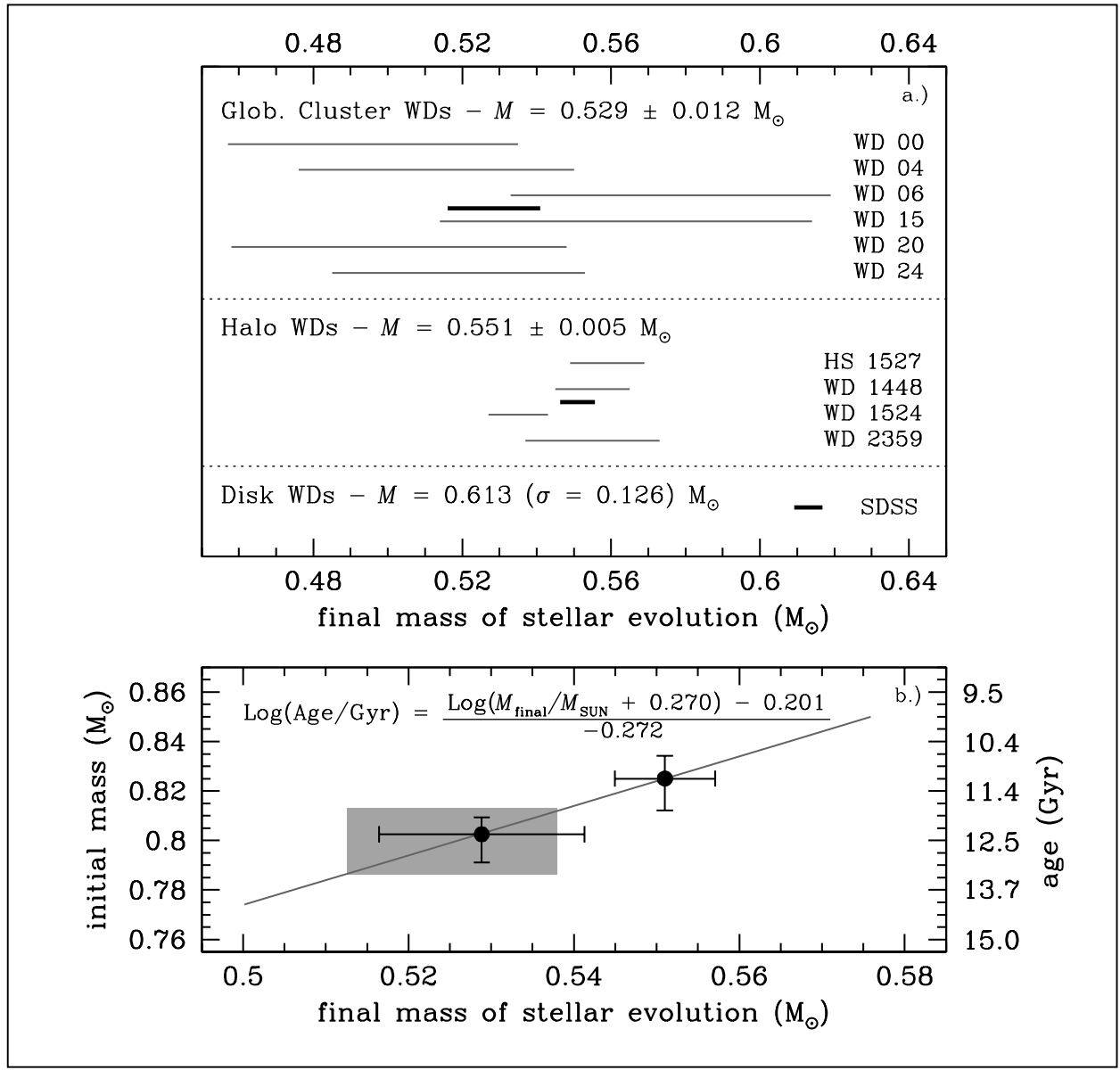

**Figure 2: The Remnant Mass and Population Age of the Milky Way Halo**

 **a.)** The masses of newly formed white dwarfs in Messier 4 (top) and the Milky Way halo (middle).  The grey bars illustrate the individual mass measurements and the black bar represents the mean of the sample.  The length of the bar indicates the uncertainty in the measurement.  The average mass of the bright Messier 4 white dwarfs is lower than the halo white dwarfs by approximately 0.02 $M_{SUN}$. **b.)** The heavier white dwarfs in the Milky Way halo formed from heavier progenitors[11,20,21]. The well-measured age of Messier 4 provides an anchor to convert the mass difference to an age difference. For a measured mass of 0.551 +/- 0.005 $M_{SUN}$ for the field halo white dwarfs, the age of the population is 11.4 +/- 0.7 billion years, similar to that of the 12.5 +/- 0.5 billion year old cluster Messier 4 given the uncertainty. The age is 2 billion years younger than the oldest globular clusters, which formed 13.5 billion years ago on this scale (the thick grey bar illustrates the age dispersion of the Galactic globular clusters)[8,23]. The mass distribution of white dwarfs in the Galactic disk from high signal-to-noise SDSS spectra, measured using the same techniques and models as described above, is $M_{FINAL}$ = 0.613 $M_{SUN}$ (1σ dispersion = 0.126 $M_{SUN}$)[28]. The distribution also has a shallow tail to higher masses (not shown). The progenitor lifetimes of these stars are much shorter than the halo and globular clusters, confirming that the Milky Way disk formed the bulk of its stars well after the halo. The newly defined relation to calculate the age of the Galactic halo cannot be directly used to date the Galactic disk for two reasons. First, the disk has had an extended star formation history and these white dwarfs are not all newly formed. Second, the progenitors of the disk white dwarfs had a very different metallicity than the halo and therefore the mass – age relation is different.